\documentclass[10pt]{article}
 
\usepackage{fullpage}
\usepackage[all=normal,bibliography=tight]{savetrees}
\usepackage{microtype}
\usepackage{todonotes}
\usepackage{amsthm}
\usepackage{amssymb,bbm}
\usepackage{microtype}
\usepackage{pdflscape}
\usepackage{multirow}

\usepackage{booktabs}

\usepackage{xspace}
\usepackage{todonotes}
\usepackage{comment}
\usepackage{afterpage}
\usepackage{pdflscape}
\usepackage{tikz}
\usepackage{graphicx}
\usetikzlibrary{arrows, decorations.markings}
\usetikzlibrary{fit}					
\usetikzlibrary{backgrounds}

\usepackage{amsmath,amssymb,amsthm}
\usepackage{graphicx}
\usepackage{caption}
\usepackage{subcaption}
\usepackage{url}
\usepackage{mathtools}

\newcommand{\Oh}{\mathcal{O}}

\numberwithin{equation}{section}

\theoremstyle{definition}

\title{Experimental Evaluation of Parameterized Algorithms for Graph Separation Problems: 
  Half-Integral Relaxations and Matroid-based Kernelization\thanks{Supported by the ``Recent trends in kernelization: theory and experimental evaluation'' project, carried out within the Homing programme of the Foundation for Polish Science co-financed by the European Union under the European Regional Development Fund.}}

\author{
  Marcin Pilipczuk\thanks{%
Institute of Informatics, University of Warsaw, Poland.
\texttt{malcin@mimuw.edu.pl}}
  \and
 Micha\l{} Ziobro\thanks{%
Theoretical Computer Science Department, Faculty of Mathematics and Computer Science, Jagiellonian University, Krak\'{o}w, Poland. \texttt{michal.18.ziobro@student.uj.edu.pl}}}

\date{}
 
\begin{document}

\maketitle

\begin{abstract}
In the recent years we have witnessed a rapid development of new algorithmic techniques for parameterized algorithms for graph separation problems. 
We present experimental evaluation of two cornerstone theoretical results in this area: linear-time branching algorithms guided by half-integral relaxations
and kernelization (preprocessing) routines based on representative sets in matroids.
A side contribution is a new set of benchmark instances of (unweighted, vertex-deletion) \textsc{Multiway Cut}.
\end{abstract}

\section{Introduction}

Since the seminal work of Marx that introduced the concept of important separators~\cite{Marx04}, the study of graph separation problems
have been an important and vivid subarea of parameterized algorithms. 
In the last decade a number of new algorithmic techniques emerged, including the aforementioned important separators and the shadow removal technique~\cite{Marx04,ChenLLOR08},
treewidth reduction~\cite{MarxOR13}, randomized contractions~\cite{ChitnisCHPP16}, preprocessing algorithms based on representative sets and gammoids~\cite{KratschW12}, and linear-time algorithms based
on half-integral relaxations~\cite{Guillemot11a,CyganPPW13,IwataWY16,IwataYY17}. The aim of this work is to provide experimental evaluation of the two last frameworks.

Recall that in parameterized complexity, an instance $x$ of a studied problem comes up with an integer parameter $k$ which is supposed to measure the complexity of the instance.
The notion of tractability, a \emph{parameterized algorithm} or an \emph{FPT algorithm}, is an algorithm that solves $(x,k)$ in time bounded by $f(k) \cdot |x|^c$ for some
computable function $f$ and a constant $c$ independent of $k$. That is, the exponential explosion in the running time, probably unavoidable for NP-hard problems, is confined
to be a function of the parameter only. If $f$ is ``reasonable'' (e.g., single-exponential) and the parameter in the given instances is small, the algorithm has chances of being practical.
In most cases in this work, we will study graph separation problems that ask to delete minimum number of vertices or edges to get some desired separation, and the parameter will be the solution size:
the minimum size of the sought solution.

\paragraph{Kernelization via representative sets.}
Kernelization, a subarea of parameterized complexity, offers a rigorous framework for analyzing preprocessing routines. 
Given an instance $x$ with a parameter $k$ of a studied parameterized problem, a \emph{kernel} is a polynomial time algorithm that reduces $(x,k)$ to an equivalent
instance $(x',k')$ with $|x'| + k' \leq g(k)$, where $g$ is a computable function called the \emph{size} of the kernel. 
On the theoretical side, the golden standard of efficiency is \emph{polynomial kernel}, where $g$ is required to be a polynomial.

A few years ago it has been a major open problem in parameterized complexity to determine whether important graph separation problems, such as \textsc{Odd Cycle Transversal} or \textsc{Multiway Cut},
admit polynomial kernels when parameterized by the solution size. 
In 2011 a breakthrough series of two papers by Kratsch and Wahlstr\"{o}m appeared~\cite{KratschW14,KratschW12} that introduced a framework of obtaining polynomial kernels via 
representative sets in linear matroids. 
By encoding the task at hand into a correct combination of gammoids and uniform matroids, Kratsch and Wahlstr\"{o}m obtained polynomial kernels for \textsc{Odd Cycle Transversal}, \textsc{Multiway Cut}
with fixed number of terminals, and a number of other graph separation problems.

The applicability of representative sets in parameterized complexity is not only limited to preprocessing, but can be also used in parameterized algorithms, for example
to limit the number of states in the dynamic programming routines in graphs of bounded treewidth. 
Experimental evaluation showed that, despite significant overhead due to involved linear algebra subroutines, if correctly fine-tuned, such approach is favorable over more direct and naive algorithms
at least for \textsc{Steiner Tree}~\cite{FafianieBN15} and \textsc{Hamiltonian Cycle}~\cite{ZiobroP18}.
These results lead to natural question: what is the efficiency of the landmark kernelization results of Kratsch and Wahlstr\"{o}m in practice?

\paragraph{Linear-time FPT algorithms via half-integral relaxations.}
There has been significant effort in the recent years on designing linear- or near-linear-time FPT algorithms; that is, parameterized algorithms where 
the $|x|^c$ factor in the running time bound is linear or near-linear in $|x|$. 
Such algorithms have biggest chances of being useful in practice in solving large instances with small parameters.

A milestone result in this direction is a work of Iwata, Wahlstr\"{o}m, and Yoshida~\cite{IwataWY16}. The authors showed that a number of graph separation problems admit half-integral relaxations
with desirable properties: they are polynomial-time solvable and persistent (some integral values of the half-integral optimum can be greedily fixed for the integral optimum). 
On one hand, by branching on values where the half-integral relaxation assigns a non-integral value one obtains parameterized algorithms with a good dependency on the parameter. 
On the other hand,~\cite{IwataWY16} showed that in most \emph{edge-deletion} cases, the half-integral relaxation can be solved in linear time. 
This mix allowed~\cite{IwataWY16} to claim very good running time bounds for a number of graph separation problems.

The first edition of Parameterized Algorithms and Computational Experiments challenge (PACE) in 2016 featured \textsc{Feedback Vertex Set} at one of its tracks~\cite{PACE2016}.
The winning entry of Imanishi and Iwata implemented the approach of Iwata, Wahlstr\"{o}m, and Yoshida~\cite{IwataWY16}, tailored for \textsc{Feedback Vertex Set}, with a custom
linear-time solver of the corresponding half-integral relaxation. (Note that \textsc{Feedback Vertex Set} is a vertex-deletion problem and thus~\cite{IwataWY16} does not provide a linear-time
algorithm solving its half-integral relaxation.) 
A subsequent work of Iwata~\cite{Iwata17} turned the solver into an improved kernelization algorithm for \textsc{Feedback Vertex Set}.
Finally, recently Iwata, Yamaguchi, and Yoshida~\cite{IwataYY17} announced a generalization of the algorithm of Imanishi and Iwata: a linear-time FPT algorithm for the so-called \textsc{CSP 0/1/all} problem that generalizes a number of graph separation
problems including \textsc{Feedback Vertex Set}, \textsc{Odd Cycle Transversal}, and \textsc{Multiway Cut}.

Following PACE 2016, Kiljan and the first author of this paper conducted an experimental study of parameterized algorithms for \textsc{Feedback Vertex Set}~\cite{KiljanP18}.
They established that the classic and based on half-integral relaxation branching algorithms are comparable in terms of efficiency, while the supremacy of the PACE 2016
entry by Imanishi and Iwata can be explained mostly by their better lower bound pruning techniques. 

This leads to the following question: as branching algorithm based on half-integral relaxations is one of the leading approaches in practice for \textsc{Feedback Vertex Set},
how does its generalization of Iwata, Yamaguchi, and Yoshida~\cite{IwataYY17} perform?
This question is particularly valid for the \textsc{Odd Cycle Transversal} problem where a number of previous experimental studies has been conducted~\cite{Wernicke03,Huffner09,AkibaI16,GHS18}.

\paragraph{Our contribution.}
In this work we address the above two questions: We perform experimental study of the representative sets-based kernelization algorithms and 
branching algorithm based on half-integral relaxations for two landmark graph separation problems, \textsc{Odd Cycle Transversal} and \textsc{Multiway Cut}.
First, we implement the parameterized algorithm for the \textsc{CSP 0/1/all} problem of Iwata, Yamaguchi, and Yoshida~\cite{IwataYY17} and provide a front-end to it for \textsc{Odd Cycle Transversal} and \textsc{Multiway Cut}.
Second, we implement the kernelization algorithms for \textsc{Odd Cycle Transversal} and \textsc{Multiway Cut} based on representative sets and matroid representations.

\paragraph{Odd Cycle Transversal.}
In the \textsc{Odd Cycle Transversal} problem, given a graph $G$ and an integer $k$, one asks for a set $X$ of at most $k$ vertices such that $G-X$ is bipartite. 
A classic result of Reed, Smith, and Vetta~\cite{ReedSV04} reduced this problem to a bounded-in-$k$ number of invocations of a minimum cut algorithm and introduced
the \emph{iterative compression} technique, now a mandatory ingredient in many modern parameterized algorithms. 
\textsc{Odd Cycle Transversal} was also the first problem to be shown to admit a polynomial kernel via matroid representation~\cite{KratschW14}. 

On the more experimental side, \textsc{Odd Cycle Transversal} has been also studied extensively. 
Wernicke~\cite{Wernicke03} in 2003 provided a first implementation of a branch-and-bound routine, later reimplemented and analyzed by H\"{u}ffner~\cite{Huffner09}.
Akiba and Iwata~\cite{AkibaI16} proved efficiency of the branching algorithm for \textsc{Vertex Cover} based on half-integral relaxation by 
showing that their solver is superior in solving \textsc{Odd Cycle Transversal} instances transformed to \textsc{Vertex Cover} via a textbook reduction.
Finally, very recently Goodrich, Horton, and Sullivan~\cite{GHS18} unified the previous experimental studies within the same source code framework
and provided a detailed comparison.

We complete the picture of Goodrich, Horton, and Sullivan~\cite{GHS18} by adding the following two ingredients to the comparison.
First, we run our implementation of the~\cite{IwataYY17} algorithm on their \textsc{Odd Cycle Transversal} benchmark.
Second, we apply our implementation of the kernelization algorithm of~\cite{KratschW12} on their \textsc{Odd Cycle Transversal} instances and compare the outcome with the reported efficiency
of simple reduction rules. 

\paragraph{Multiway Cut.}
In the \textsc{Multiway Cut} problem, given a graph $G$, a set of terminals $T \subseteq V(G)$, and an integer $k$, one seeks for a set $X \subseteq V(G) \setminus T$ of size at most $k$
such that every connected component of $G-X$ contains at most one terminal. 
Fixed-parameter tractability of \textsc{Multiway Cut} parameterized by the solution size has been established by Marx~\cite{Marx04} in the work that introduced the seminal notion of important 
separators. 
A classic algorithm based on important separators is known to run in time $4^k k^{\Oh(1)} (n+m)$~\cite{ChenLLOR08} (see~\cite{thebook} for a textbook exposition).
A branching algorithm based on half-integral relaxations has theoretical running time bound of $2^k k^{\Oh(1)} (n+m)$~\cite{CyganPPW13,IwataYY17}.

The representative sets framework provides a kernel for \textsc{Multiway Cut} of size $\Oh(k^{|T|+1})$, that is, polynomial for a fixed number of terminals. 
It is a major open problem in the are to determine if the exponential dependency on $|T|$ is necessary in the size of the kernel.

To the best of our knowledge, no prior experimental evaluation of parameterized algorithms for \textsc{Multiway Cut} has been performed.
In particular, there is no established benchmark set of instances as it is the case for \textsc{Odd Cycle Transversal}. 
Our first contribution is providing such a set. Our instances come from three sources: artificially hard instances coming from an NP-hardness reduction from \textsc{Max Cut}, 
random instances with planted solution, and sparse graphs with randomly chosen set of terminals. We discuss our benchmark set in detail in Section~\ref{ss:mwc:tests}.

We implement the classic algorithm based on important separators~\cite{Marx04,ChenLLOR08} and compare it with the algorithm for \textsc{CSP 0/1/all}~\cite{IwataYY17} on our benchmark set. 
Furthermore, we apply the kernelization algorithm on a small subset of simplest tests in our benchmark. 

\medskip

We were very recently informed that Fafianie and Kratsch are also independently conducting experiments on matroid-based kernelization of \textsc{Multiway Cut}~\cite{stefans}.

\paragraph{Organization.}
We give a more detailed overview of the studied algorithms in Section~\ref{sec:algs}.
In Section~\ref{sec:setup} we explain the setup for the experiments, in particular, we discuss the benchmark set of instances.
Results are presented in Section~\ref{sec:results}.
Section~\ref{sec:conc} concludes the paper.

\section{Overview of the tested algorithms}\label{sec:algs}

\subsection{Branching algorithms via half-integral relaxations}

The input to a binary CSP problem
consists of a \emph{domain} $D$ and a \emph{constraint graph} $G$.
Each vertex $v \in V(G)$ corresponds to a variable (which we will also denote by $v$)
and each edge $uv \in E(G)$ corresponds to a constraint $\psi_{u,v} \subseteq D \times D$.
An assignment $\phi : V(G) \to D$ \emph{satisfies} a constraint $\psi_{u,v}$ if
$(\phi(u), \phi(v)) \in \psi_{u,v}$.
Somewhat abusing the notation, we assume that if a constraint $\psi_{u,v}$ is present,
         then also a reversed constraint $\psi_{v,u}$ is present.

In the \textsc{CSP 0/1/all} problem, each constraint $\psi_{u,v}$ is required to satisfy
the following ``0/1/all'' property: for every $a \in D$ the number of possible values $b \in B$
such that $(a,b) \in \psi_{u,v}$ is either $0$, $1$, or $|D|$, and a symmetric condition
holds if we fix $b \in D$ and ask for the number of $a \in D$ with $(a,b) \in \psi_{a,b}$.
That is, fixing one variable in a constraint either makes it immediately unsatisfied, makes
it immediately satisfied, or forces the value of the second variable.

There are two important examples of 0/1/all constraints. First, observe that any permutation
of $D$ is satisfies the 0/1/all property. Second, the ``2-SAT'' constraint $(u = a) \vee (v = b)$,
that is satisfied if $\phi(u) = a$ or if $\phi(v) = b$, satisfies the 0/1/all property.

The \textsc{CSP 0/1/all} problem, given a binary CSP instance with all constraints
satisfying the 0/1/all property and an integer $k$, asks to delete at most $k$ variables from
the instance so that the remaining part admits an assignment satisfying all constraints.
We also
consider an ``extension'' variant of the problem where we are additionally given a partial assignment
$\phi_0 : A \to D$ for some $A \subseteq V(G)$ and we ask to delete a set $X$ of at most $k$
variables such that $\phi_0|_{A \setminus X}$ can be extended to $V(G) \setminus X$ while
satisfying all constraints. 

The \textsc{Odd Cycle Transversal} problem can be cast as a \textsc{CSP 0/1/all} problem
by defining $|D| = 2$ and all edges of $G$ being constraints ``not equal''
(i.e., $(a,b) \in \psi_{u,v}$ if $a \neq b$). 
A \textsc{Multiway Cut} instance $(G,T,k)$ can be cast as a \textsc{CSP 0/1/all Extension} instance
by the following steps:
\begin{itemize}
\item Simple preprocessing: reject the instance if there is an edge connecting two terminals,
  and assign to the sought solution (i.e., delete from $G$ and decrease $k$ accordingly)
  all vertices that are adjacent to at least two terminals.
\item Define $D = T$, $A = N(T)$, and $\phi_0(v) = t$ for every $v \in N(t)$ and $t \in T$.
\item Output $(D,G-T,k,\phi_0)$ where every constraint is the 
equality constraint $\psi_{u,v} = \{(t,t)~|~t \in T\}$.
\end{itemize}

The nature of 0/1/all constraints allows to define \emph{implicational paths} as sequences
of pairs $(v_i,a_i)_{i=1}^\ell$, $v_i \in V(G)$, $a_i \in D$, such that:
\begin{itemize}
\item $v_1 \in A$ and $\phi_0(v_1) = a_1$;
\item for every $2 \leq i \leq \ell$ there exists a constraint $\psi_{v_{i-1},v_i}$
such that $a_i$ is the only value with $(a_{i-1},a_i) \in \psi_{v_{i-1},v_i}$.
\end{itemize}
An implicational path $(v_i,a_i)_{i=1}^\ell$ is \emph{conflicting} if $v_\ell \in A$
but $\phi_0(v_\ell) \neq a_\ell$.
Note that any solution to \textsc{CSP 0/1/all Extension} on $(D,G,k,\phi_0)$
needs to delete at least one vertex of each conflicting path.

An \emph{integral packing} of conflicting paths is a set of vertex-disjoint conflicting paths.
A \emph{half-integral packing} of conflicting paths is a set of conflicting paths such that
every vertex of $G$ lies on at most two paths. 
A \emph{slack} of a \textsc{CSP 0/1/all Extension} is defined as
$k - |\mathcal{F}|/2$ where $\mathcal{F}$ is
maximum-size half-integral packing of conflicting paths.
Note that if $\mathcal{F}$ is a half-integral packing of conflicting paths, then every solution
(deletion set)
to \textsc{CSP 0/1/all Extension} has size at least $|\mathcal{F}|/2$, and thus
slack of a yes-instance is nonnegative.

For a function $f : V(G) \to \mathbb{R}$, a \emph{weight} of a set $Z \subseteq V(G)$
is defined as $f(Z) := \sum_{v \in Z} f(v)$. The weight of an implicational path
$P = (v_i,a_i)_{i=1}^\ell$ is defined as $\sum_{i=1}^\ell f(v_i)$. 
A \emph{half-integral cover} in an instance $(D,G,k,\phi_0)$ is an assignment
$f : V(G) \to \{0,1/2,1\}$ such that for every conflicting path $(v_i,a_i)_{i=1}^\ell$
has weight at least one.
Note that minimum weight of a half-integral cover is a lower bound for a solution
size to \textsc{CSP 0/1/all Extension}.

Given an instance $(D,G,k,\phi_0)$ of \textsc{CSP 0/1/all Extension}, we define
two basic operations on it. First, we can proclaim a vertex $v \in V(G)$ to be in the
sought solution.  That involves deleting $v$ (from $G$ and from the domain of $\phi_0$
if it is there) and decreasing $k$ by one. 
Second, we can proclaim a vertex $v \in V(G)$ with $\phi_0(v)$ defined undeletable.
To proclaim $v$ undeletable, we inspect every constraint $\psi_{v,u}$. If 
$\phi_0(u)$ is defined and $(\phi_0(v),\phi_0(u)) \in \psi_{v,u}$ or every $b \in D$
satisfies $(\phi_0(v),b) \in \psi_{v,u}$, we do nothing for the constraint $\psi_{v,u}$.
If $\phi_0(u)$ is defined and $(\phi_0(v),\phi_0(u)) \notin \psi_{v,u}$ or 
there is no $b \in D$ with $(\phi_0(v),b) \in \psi_{v,u}$, then we proclaim
$u$ to be in the sought solution.
In the remaining case, if $\phi_0(u)$ is undefined and there is exactly one $b \in D$
such that $(\phi_0(v), b) \in \psi_{v,u}$, then we set $\phi_0(u) = b$.
Finally, after all constraints $\psi_{v,u}$ have been inspected, we 
delete $v$ from $G$.

The main technical contribution of Iwata, Yamaguchi, and Yoshida~\cite{IwataYY17}
is the following:
\begin{enumerate}
\item Given a \textsc{CSP 0/1/all Extension} instance $(D,G,k,\phi_0)$ one can
in time $\Oh(Tk|G|)$ compute a maximum-size half-integral packing $\mathcal{F}$ 
of conflicting paths. 
Here, $T$ stands for time needed for a number of simple operations on the constraints
(which constant or near-constant in all applications relevant for this paper).
The algorithm is an involved augmenting-path-like algorithm that iteratively extends the
half-integral packing until possible.
\item Given the final state of the augmenting-path algorithm, one can in linear time obtain
a minimum-weight half-integral cover $f$
with the additional ``furthest cover'' and persistence 
properties:
\begin{enumerate}
\item For every implicational path $P=(v_i,a_i)_{i=1}^\ell$ of weight zero,
  one can iteratively proclaim all vertices $v_1,v_2,\ldots,v_\ell$ undeletable.
\item If $(D,G,k,\phi_0)$ is a yes-instance, then there exists a solution $X$ with
$f^{-1}(1) \subseteq X$ (and thus one can greedily proclaim all vertices of $f^{-1}(1)$ to be in
    the sought solution).
\item Once the above preprocessing steps have been performed, branching on a vertex $v \in f^{-1}(1/2)$ leads to a progress in the following sense. 
First, it will always be the case that $\phi_0(v)$ is defined.
In one subcase, we proclaim $v$ to be in the sought solution.
Then the maximum size of a half-integral packing of conflicting paths drops by only one,
     and thus the slack decreases by a half.
In the other case, we proclaim $v$ to be undeletable, and this step increases the size of a maximum half-integral packing of conflicting paths by at least one, thus decreasing slack by at least
a half.
\end{enumerate}
\end{enumerate}
Iteratively applying the above steps (and branching on a value of an arbitrary vertex
    if $\phi_0 = \emptyset$) leads to an algorithm with running time bound
$\Oh(|D|^{2k} \cdot k \cdot |G|)$ for \textsc{CSP 0/1/all Extension}.
For \textsc{Odd Cycle Transversal}, this becomes $\Oh(4^k k |G|)$.
For \textsc{Multiway Cut}, one can observe that the first half-integral packing of
conflicting paths computed by the algorithm is of size at least $k$ and thus
the running time bound becomes $\Oh(2^k \cdot k \cdot |G|)$. 

The implementation of the algorithm if Iwata, Yamaguchi, and Yoshida~\cite{IwataYY17}
is rather direct, without any significant simplifications nor omissions. 
It is available in the \texttt{csp-0-1-all} subdirectory of the code repository~\cite{repo}.

Both for \textsc{Odd Cycle Transversal} and \textsc{Multiway Cut}, we implemented a preprocessing routine that
invokes the algorithm up to just before the first branching step and outputs the reduced instance.
Furthermore, for both problems the algorithm is run in four variants.
The first, denoted later by \texttt{CSP}, is the basic variant that does not take into account the size of the computed maximum half-integral packing as a solution
lower bound. 
The variant \texttt{CSP LB1} takes it into account and prunes the search tree if the slack becomes negative.

The variant \texttt{CSP LB2} is a slight improvement of the previous variant: in the algorithm of~\cite{IwataYY17} one can deduce from the special structure
of the constructed half-integral packing a slightly higher lower bound than merely half of the number of paths. 
In short, the constructed half-integral packing is a disjoint union of paths and ``flowers'' which are unions of conflicting paths arranged in a special fashion.
Size of an integral solution to \textsc{CSP 0/1/all Extension} is lower bounded by half of the total number of packed paths and the number of flowers in the packing.%
\footnote{We thank Yoichi Iwata for showing us this optimization.}

The variant \texttt{CSP+red} is a pipe-lined aforementioned preprocessing and, as a separate call, the algorithm \texttt{CSP}. It is meant only for checking
correctness, as expected is slightly slower than merely running \texttt{CSP}, and will not be discussed later.

Furthermore, for \textsc{Multiway Cut}, each of the aforementioned variants is also run in a \texttt{+CC} variant where, in case of a disconnected graph, each connected component is solved
independently.

As discussed in the introduction, we compared the \textsc{CSP 0/1/all Extension} solver for \textsc{Multiway Cut} with the classic approach via important separators~\cite{Marx04}; see~\cite{thebook}
for a textbook exposition of the algorithm. This algorithm also comes with the \texttt{+CC} connected components variant.
In later discussion and the \texttt{results-MWC.csv} file in the repository~\cite{repo} this algorithm is denoted by \texttt{ImpSeps}.

\subsection{Kernelization via representative sets}

The main engine of the studied kernelization algorithms lies in the notion of 
\emph{representative sets}, defined by Lov\'{a}sz~\cite{Lovasz77}
and introduced to the parameterized complexity
setting by Marx~\cite{Marx09}. 
Let $V$ be a finite set of vectors in $\mathbb{F}^n$ spanning a subspace of dimension $r$
(i.e., the vectors with the notion of linear independence form a linear matroid over $\mathbb{F}$
 of rank $r$). 
Let $\mathcal{F} \subseteq \binom{V}{a}$ be a family of linearly independent $a$-tuples of
vectors from $V$ (i.e., every $a$-tuple is a set of $a$ linearly independent vectors)
and let $b = r-a$. 
We say that $\mathcal{F}^\ast \subseteq \mathcal{F}$ is a \emph{$a$-representative set} 
if for every $B \in \binom{V}{b}$, if there exists $A \in \mathcal{F}$ such that $A \cap B = \emptyset$
and $A \cup B$ is a linearly independent set of vectors (of size $a + b = r$), then
there exists $A^\ast \in \mathcal{F}^\ast$ with the same properties: $A^\ast \cap B = \emptyset$
and $A^\ast \cup B$ are linearly independent.

The main result of Lov\'{a}sz is that there exists a $a$-representative set of size at most
$\binom{r}{a}$ and it can be computed by associating to every $A \in \mathcal{F}$
a carefully chosen vector $w_A \in \mathbb{F}^{\binom{r}{a}}$ and then outputting as $\mathcal{F}^\ast$ sets $A$ that correspond to a maximal linearly independent set of vectors 
from $\{w_A~|~A \in \mathcal{F}\}$. 

On high level, the kernelization algorithm for \textsc{Odd Cycle Transversal} computes
an approximate solution $Y$, constructs the aforementioned set of vectors $V$ based
on a properly defined gammoid in a slightly modified graph $G$, constructs for every $v \in V(G)$
a triple $A_v \in \binom{V}{3}$, and finds a $3$-representative set $\mathcal{F}^\ast \subseteq 
\{A_v~|~v \in V(G)\}$. The essence of the construction lies in the following property: for every $x \notin Z$
there there exists an optimal solution to the input \textsc{Odd Cycle Transversal} instance that does not contain $x$.
Thus, we can iteratively proclaim such $x$ undeletable and repeat.
The rank of $V$ is $\Oh(|Y|)$, and thus we obtain $|Z| = \Oh(|Y|^3)$ as the final size of the kernel.

Our implementation constructs an approximate solution $Y$ greedily and then computes the set $Z$
according to the above description.

For \textsc{Multiway Cut}, the kernelization algorithm of Kratsch and Wahlstr\"{o}m~\cite{KratschW12}
goes along similar lines. The constructed set of vectors $V$ is of rank $\Oh(|N(T)|+k)$. 
We look for $(|T|+1)$-representative set. The final set $Z$ is not guaranteed to contain
an optimal solution; instead, we know only that for every $v \notin Z$ there exists an optimal
solution not containing $Z$. Thus we can proclaim such $v$ undeletable (by turning its neighborhood into a clique and deleting $v$) and restart the algorithm.

In the original work~\cite{KratschW12} one obtains $|N(T)| \leq 2k$ by a well-known preprocessing
routines involving solving a linear relaxation of the problem. 
This gives a $\Oh(k^{|T|+1})$ bound on the output set $Z$.
We observe that the same property $|N(T)| \leq 2k$ can be obtained by applying the first step
of the \textsc{CSP 0/1/all Extension} algorithm before the first branching step. 
Thus, we start our kernelization algorithm by first running the algorithm described in the previous section and stopping it just before the first branching step.

Furthermore, to simplify the task of the kernelization algorithm, we feed it with the optimal
value $k$ of the solution size.

\section{Setup}\label{sec:setup}

\subsection{Hardware and software}

The experiments with the \textsc{CSP 0/1/all Extension} solver
have been performed on a cluster of
16 computers at the Institute of Informatics, University of Warsaw.
Each machine was equipped with Intel\textsuperscript{\textregistered} Xeon\textsuperscript{\textregistered} E3-1240v6 3.70GHz processor
and 16 GB RAM.
All machines shared the same NFS drive. Since the size of the inputs and
outputs to the programs is small, the network communication
was negligible during the process.

Computations of the kernel for \textsc{Odd Cycle Transversal}
were performed on the computer of Jagiellonian University
with Intel\textsuperscript{\textregistered} Xeon\textsuperscript{\textregistered} Gold 6154 3.00GHz processors
and about 1 TB of random access memory, operated by Ubuntu.

Computations of the kernel for \textsc{Multiway Cut} were performed on a PC
with an Intel\textsuperscript{\textregistered} Core i7-6700 processor and 32 GB of random-access memory.
The operating system used during the experiments was Ubuntu 18.04.

The code has been written in C++ and is available at~\cite{repo}.
To manage a field modulo some large prime in the kernelization algorithms
we used GNU Multiprecision Arithmetic Library.

\subsection{Tests for Odd Cycle Transversal}

In terms of data sets for \textsc{Odd Cycle Transversal} we use an established
benchmark set from~\cite{GHS18}.
It consists of the well known set provided by Wernicke \cite{Wernicke03}
along with the unweighted versions of small instances from the Beasley set~\cite{beasley1998heuristic} (50 and 100 vertices) and the GKA~\cite{gka} set, both of which are originally instances of the QUBO problem.
We refer to~\cite{GHS18} for more discussion on this benchmark set of instances.

As in~\cite{GHS18}, we removed weights, loops and multiple edges from all of the instances.
The prepared instances are written into files organized as follows:
In the first line we give the number of vertices and edges and
in the next lines (one per edge) we put two integers corresponding
to vertices connected with one of the edges.

For sake of simplicity, we use the same names for instances same as in \cite{GHS18}.

\subsection{Tests for Multiway Cut}\label{ss:mwc:tests}

As discussed in the introduction, we are not aware of any established benchmark
set of instances for \textsc{Multiway Cut}. 
We constructed our own set of instances coming from three sources.
In the repository, tests are gathered in the \texttt{mwc-tests} subdirectory;
each test has an extension \texttt{.mwc} and a corresponding
\texttt{.opt} file contains an optimum solution.
A CSV file \texttt{results-mwc.csv} contains basic statistics of all tests.

\paragraph{NP-hardness reduction from \textsc{Max Cut}.}
A classic reduction of NP-hardness of \textsc{Multiway Cut} with three terminals
due to Dahlhaus et al.~\cite{DahlhausJPSY94} starts from the \textsc{Max Cut} problem. The crux of the reduction
is a constant-size gadget that exhibits non-submodularity of the solution space.
Originally the reduction leads to an edge-deletion variant of \textsc{Multiway Cut} with edges
of weight $1$ or $4$, but it is straightforward to adapt it to unweighted vertex-deletion
\textsc{Multiway Cut}. 
All instances obtained in this manner have three terminals.

Our first set of benchmark instances comes from this reduction, applied to a number of constant-sized
graphs: $K_2$, $K_3$, $K_4$, $C_4$, $C_5$, $P_3$ (a path on three vertices), 
a bull (a triangle with two degree-one vertices adjacent to two distinct vertices of the triangle),
and a pan (a $C_4$ with an extra degree-one vertex adjacent to one of the vertices of the cycle). 
Their names are \texttt{maxcut\_G} where $G$ is the source graph for the reduction.

\paragraph{Planted solution.}
Let $t$, $a$, and $b$ be integers and let $0 \leq p \leq 1$ be a real.
Let $H(t,a,b)$ be a graph consisting of a clique $K_0$ on $a$ vertices, $t$ cliques
$K_1,K_2,\ldots,K_t$ on $b$ vertices each, and terminals $\{t_i~|~1 \leq i \leq t\}$,
with the following adjacencies: each clique $K_i$ for $1 \leq i \leq t$ is fully adjacent
to $K_0$ and to $t_i$.
If $a < (t-1)b$ then $K_0$ is the unique minimum solution to \textsc{Multiway Cut} on this instance. 
Let $G(t,a,b,p)$ be a subgraph of $H(t,a,b)$ with the same vertex set and each edge
of $H(t,a,b)$ present in $G(t,a,b,p)$ independently with probability $p$.
For a good choice of $t$, $a$, $b$, and $p$, the set $K_0$ still remains a unique
minimum solution to \textsc{Multiway Cut} on $G(t,a,b,p)$, but the random choice of edges
makes it more ``fuzzy''.

We generated a number of graphs $G(t,a,b,p)$ for $t \in \{3,4,5,7,10\}$, 
$a,b \in \{50,100,200\}$, and $p \in \{0.2,0.3,0.4\}$.
Their file names in the repository are of the form \texttt{planted\_a\_t\_b\_p}
where $p$ is expressed in percents.
On each computed graph we run a simple approximation algorithm that iteratively separates
a terminal from the remaining terminals by a minimum cut and kept only these instances
where this approximation algorithm returned a suboptimal result.

\paragraph{Sparse graphs.}
We generated also a number of sparse \textsc{Multiway Cut} instances as follows.
We took a number of sparse graphs from established benchmarks~\cite{Kunegis13,NadaraPRRS18}
consisting
of real-world graphs and random planar graphs.
In each such graph $G$, for every $t \in \{3,4,5,\mathrm{max}\}$, we randomly chose
a $5$-scattered set $S$ (i.e., every two vertices of $S$ are within distance at least $5$; $t = \mathrm{max}$ stands for a maximal $5$-scattered set), contracted $N(s)$ onto $s$ for every $s \in S$
and proclaimed $S$ to be the terminal set in the resulting graph. 
As before, from all instances we kept only these where the aforementioned approximation algorithm
returned a suboptimal solution. 

The filenames of these tests are \texttt{sparse\_name\_t} where \texttt{name} is the original
name of the test from~\cite{NadaraPRRS18} and $t$ is the size of the found $5$-scattered set 
(here $0$ stands for $\mathrm{max}$). 

\section{Results}\label{sec:results}

\subsection{Branching algorithms}

\subsubsection{Odd Cycle Transversal}

\begin{table}[tbh]
\begin{center}
\begin{tabular}{l|lll|lll}
test & $|V(G)|$ & $|E(G)|$ & OPT & \texttt{CSP} & \texttt{CSP LB1} & \texttt{CSP LB2} \\\hline
\texttt{aa10} & 69 & 191 & 6 & 0:00.11 & 0:00.04 & 0:00.04 \\
\texttt{aa11} & 103 & 307 & 11 & 0:00.86 & 0:00.26 & 0:00.25 \\
\texttt{aa13} & 131 & 383 & 12 & 0:02.97 & 0:00.91 & 0:00.89 \\
\texttt{aa14} & 127 & 525 & 19 & 0:38.53 & 0:05.56 & 0:05.44 \\
\texttt{aa15} & 67 & 179 & 7 & 0:00.15 & 0:00.05 & 0:00.05 \\
\texttt{aa17} & 153 & 633 & 25 & 19:07.83 & 2:21.74 & 2:18.55 \\
\texttt{aa18} & 90 & 381 & 14 & 0:05.80 & 0:01.82 & 0:01.82 \\
\texttt{aa19} & 192 & 645 & 19 & 6:34.78 & 0:56.69 & 0:55.11 \\
\texttt{aa20} & 225 & 766 & 19 & 21:52.32 & 3:38.99 & 3:29.90 \\\hline
\texttt{bqp50\_1} & 50 & 108 & 11 & 0:02.25 & 0:00.23 & 0:00.30 \\
\texttt{bqp50\_2} & 50 & 120 & 11 & 0:00.53 & 0:00.07 & 0:00.06 \\
\texttt{bqp50\_3} & 50 & 132 & 14 & 0:06.12 & 0:00.44 & 0:00.41 \\
\texttt{bqp50\_4} & 50 & 111 & 11 & 0:00.46 & 0:00.08 & 0:00.07 \\
\texttt{bqp50\_5} & 50 & 131 & 13 & 0:03.77 & 0:00.42 & 0:00.41 \\
\texttt{bqp50\_6} & 50 & 101 & 9 & 0:00.17 & 0:00.04 & 0:00.03 \\
\texttt{bqp50\_7} & 50 & 124 & 10 & 0:01.18 & 0:00.12 & 0:00.11 \\
\texttt{bqp50\_8} & 50 & 137 & 14 & 0:08.37 & 0:00.82 & 0:00.81 \\
\texttt{bqp50\_9} & 50 & 122 & 12 & 0:02.64 & 0:00.24 & 0:00.23 \\
\texttt{bqp50\_10} & 50 & 108 & 11 & 0:01.94 & 0:00.25 & 0:00.23 \\\hline
\texttt{gka\_20} & 50 & 763 & 39 & 0:00.80 & 0:00.39 & 0:00.38 \\
\texttt{gka\_21} & 60 & 701 & 40 & 0:43.13 & 0:06.67 & 0:06.44 \\
\texttt{gka\_22} & 70 & 720 & 43 & 20:31.66 & 1:34.42 & 1:29.96 \\
\texttt{gka\_31} & 100 & 2948 & 85 & 3:14.24 & 1:14.49 & 1:12.37 \\
\texttt{gka\_32} & 100 & 3434 & 88 & 0:32.97 & 0:17.00 & 0:16.81 \\\hline
\end{tabular}
\caption{Selected results for \textsc{Odd Cycle Transversal}.}\label{tb:oct}
\end{center}
\end{table}

For \textsc{Odd Cycle Transversal}, we gave a 30 minutes time limit on each instance.
All running times of all algorithms, together with basic statistics of the tests, can be found in \texttt{results-OCT.csv} file in the repository~\cite{repo}.
In Table~\ref{tb:oct} we present results for a selected subset of the tests.

We observe that our reported running times are much worse than the ones
reported by Goodrich, Horton, and Sullivan~\cite{GHS18} for other approaches.
The differences, in particular when compared with the solver of Akiba and Iwata~\cite{AkibaI16},
are of the order of magnitudes and cannot be explained neither by the difference in the hardware
nor low-level implementation details. 

It is worth noting that the algorithm of Akiba and Iwata~\cite{AkibaI16} operates on the same
principles (half-integral relaxations) as the \textsc{CSP 0/1/all Extension} algorithm studied
in this paper. However, their approach is closely tailored for \textsc{Vertex Cover}
which is an explanation for their much better performance.

We conclude that the generic algorithm for \textsc{CSP 0/1/all Extension} is no match in 
practice with the tailored solutions for \textsc{Odd Cycle Transversal}.

\subsubsection{Multiway Cut}

Full results of experiments for \textsc{Multiway Cut} can be found in the \texttt{results-MWC.csv} file in the repository~\cite{repo}. Each test was run with time limit of 5 hours.
Apart from the running times and test names, the file provides the following information:
\begin{description}
\item[$n$, $m$, $t$] the number of vertices, edges, and terminals, respectively;
\item[LB1, LB2] the initial value of the two lower bounds stemming from the half-integral packing of conflicting paths, used by the algorithms \texttt{CSP LB1} and \texttt{CSP LB2}, respectively;
\item[OPT] optimum size of the solution
\item[APPX] size of an approximate solution computed by the algorithm that iteratively separates a terminal with a minimum cut from the remaining terminals;
\item[PRE DEL, PRE UNDEL, PRE RES TERMS] the number of vertices assigned by the preprocessing step
to solution, as undeletable, and as already resolved terminals (in their own connected component),
   respectively;
\item[PRE LEFT TERM, PRE LEFT DEG] the number of remaining terminals and neighbors of terminals after preprocessing;
\item[PRE LEFT OPT] optimum value of the preprocessed instance.
\end{description}
Subdirectory \texttt{mwc-tests/preprocessed} contains output instances from the preprocessing step.

\paragraph{Preprocessing.}
We first discuss the results of preprocessing step based on the half-integral packing
computed by the algorithm. That is, we run the algorithm for \textsc{CSP 0/1/all Extension}
and stop it just before the first branching step, but after all the reductions performed.

As could be expected, the preprocessing step is not able to make any reductions in the instances
coming from the NP-hardness reduction from \textsc{Max Cut}; basic statistics for these tests
are presented in Table~\ref{tb:pre-mwc:maxcut}.
The situation changes dramatically for other groups of tests.
Out of 86 planted tests, the preprocessing step completely resolved 65, and in the remaining
21 instances reduced the size significantly (around a half). 
Out of 23 sparse tests, only one test has been resolved completely by the preprocessing step,
but in most tests the preprocessing step reduced more than half of the instance.
The results for planted and sparse instances are presented in Tables~\ref{tb:pre-mwc:planted}
and~\ref{tb:pre-mwc:sparse}, respectively.

\begin{table}[tbh]
\begin{center}
\begin{tabular}{l|lll|llll}
test & $|V(G)|$ & $|E(G)|$ & $t$ & LB1 & LB2 & OPT & APPX \\\hline
\texttt{maxcut\_bull} & 273 & 2008 & 3 & 120 & 120 & 136 & 140 \\
\texttt{maxcut\_c4} & 219 & 1504 & 3 & 96 & 96 & 108 & 112 \\
\texttt{maxcut\_c5} & 273 & 1880 & 3 & 120 & 120 & 136 & 140 \\
\texttt{maxcut\_k2} & 57 & 312 & 3 & 24 & 24 & 27 & 28 \\
\texttt{maxcut\_k3} & 165 & 1128 & 3 & 72 & 72 & 82 & 84 \\
\texttt{maxcut\_k4} & 327 & 2640 & 3 & 144 & 144 & 164 & 168 \\
\texttt{maxcut\_p3} & 111 & 688 & 3 & 48 & 48 & 54 & 56 \\
\texttt{maxcut\_pan} & 273 & 1944 & 3 & 120 & 120 & 135 & 140 \\
\end{tabular}
\caption{Basic statistics for instances coming from \textsc{Max Cut}.}\label{tb:pre-mwc:maxcut}
\end{center}
\end{table}

\paragraph{Running times.}
Table~\ref{tb:mwc} gathers running times for selected instances, omitting ones that are completely resolved by the preprocessing routine
and ones with negligible (below half a second) running time of all algorithms. 

First, observe that splitting into connected components is critical for the performance of the classic approach via important separators (\texttt{ImpSeps}),
while it is not essential for the variants of the algorithm for \textsc{CSP 0/1/all Extension} (and often decreases performance). 
Second, the lower bound pruning is critical for the performance of the \texttt{CSP} variants; the second stronger lower bound increases the performance further
by a 10-20\% margin on \textsc{Max Cut} instances. 

Third, and most importantly, the best variant of the \textsc{CSP 0/1/all Extension} algorithm, namely \texttt{CSP LB2}, clearly outperforms the classic
approach \texttt{ImpSep+CC} on all planted and sparse tests. On the other hand, for the artificial instances coming from the NP-hardness reduction from \textsc{Max Cut},
\texttt{ImpSep+CC} clearly outperforms other approaches. This can be explained by a relatively short distances between terminals in this set of instances
allowing the \texttt{ImpSep+CC} algorithm to effectively search the solution space for the underlying \textsc{Max Cut} instance (which is very small).

\subsection{Kernelization}

In short, the obtained results for matroid-based kernelization indicates that this approach is impractical.

\subsubsection{Odd Cycle Transversal}

Comparing to preprocessing by Goodrich et al.~\cite{GHS18} our implementation performs worse on every instance of Wernicke's dataset.
Moreover, on every instance their algorithm removes at least as many vertices as ours returns as not necessary for the optimal OCT solution.
While the discussed approach returns as undeletable $6.75\%$ of vertices on average, the one by Goodrich et al.~\cite{GHS18} removes $29.24\%$ of vertices, marks as undeletable $29.24\%$ of vertices and as present in some optimal solution $1.60\%$ of vertices on average.
In Table~\ref{tb:oct_res} we compare the results of both algorithms on a few chosen instances.

\begin{table}[bth]
\begin{center}
\begin{tabular}{l|rr|r|rrr}
Test name & $|V|$ & $|E|$ & $|U|$ & $|B_r|$ & $|B_o|$ & $|B_i|$  \\\hline
 jap-18.gr & 73 & 296 & 12 & 15 & 0 & 11\\
 aa-23.gr & 143 & 508 & 16 & 21 & 0 & 69\\
 aa-30.gr & 39 & 71 & 5 & 9 & 0 & 12\\
 aa-22.gr & 167 & 641 & 0 & 5 & 0 & 115\\
 aa-48.gr & 90 & 343 & 0 & 14 & 0 & 35\\
 aa-47.gr & 65 & 229 & 13 & 17 & 1 & 13\\
\hline\end{tabular}
\caption{Comparison of the performance on a few sample instances. $U$ denotes the set returned by our algorithm and $B_r$, $B_o$, $B_i$
			 denote the sets of vertices (accordingly) removed, marked as necessary and marked as undeletable by the algorithm of Goodrich et al.~\cite{GHS18}.}\label{tb:oct_res}
\end{center}
\end{table}

On the other hand, we can observe that, in a few cases, we get a result much better than the theoretical bound of $8|Y|^3$, where $Y$ is a given approximate solution.
which could be caused by a nontrivial internal structure of the considered matroid.
As the instances are rather small and their optimal solutions are relatively big, such behavior can be observed on almost all instances for which
we found a nonempty set of undeletable vertices. We present some of them in Table~\ref{tb:innerstruct}.

\begin{table}[bth]
\begin{center}
\begin{tabular}{l|rr|rr|rr}
Test name & $|V|$ & $|E|$ & $|Y|$ & $8|Y|^3$ & $|U|$ & $|V \setminus U|$\\\hline
 jap-24.gr & 150 & 387 & 4 & 512 & 46 & 104\\
 jap-28.gr & 96 & 567 & 21 & 74088 & 15 & 81\\
 jap-23.gr & 76 & 369 & 23 & 97336 & 21 & 55\\
 aa-29.gr & 277 & 1058 & 29 & 195112 & 10 & 267\\
 aa-23.gr & 143 & 508 & 21 & 74088 & 16 & 127 \\
 aa-35.gr & 82 & 269 & 13 & 17576 & 5 & 77\\
\hline\end{tabular}
\caption{Results on some instances in comparison to the theoretical bound on the size of the resulting set of deletable vertices shown in~\cite{KratschW12},
			where $Y$ is some given approximate solution and $U$ is the result set of undeletable vertices (so $V \setminus U$ is the set of vertices left as deletable).}\label{tb:innerstruct}
\end{center}
\end{table}

Our implementation could not find a single undeletable vertex on any instance from GKA and Beasley datasets.
This is not surprising, as the theoretical bound on the resulting set of deletable vertices
has a cubic dependence on the size of
a given solution and, according to~\cite{GHS18},
the optimal solutions of these instances are quite large compared to the sizes of the instances.
For example, for larger Beasley instances (i.e.\ those with $100$ vertices) the optimal solutions contain approximately half of their vertices,
while in case of a part of the GKA instances we need to remove almost all of the vertices to make them bipartite.

\subsubsection{Multiway Cut}

With the rather poor performance of the kernelization algorithm for \textsc{Odd Cycle Transversal}, we
decided to make the task of the \textsc{Multiway Cut} kernel as easy as possible.

We applied it to all instances of \textsc{Multiway Cut} after the preprocessing step described in Tables~\ref{tb:pre-mwc:planted} and~\ref{tb:pre-mwc:sparse}. 
As mentioned earlier, the evaluation has been performed on a PC with 32 GB of RAM. Only on one instance, \texttt{sparse\_planar400\_0}, the algorithm did not run out
of memory and proclaimed 13 additional vertices as undeletable.

\section{Conclusions}\label{sec:conc}

We have presented an experimental evaluation of two cornerstone theoretical results in parameterized complexity of graph separation problems:
branching algorithms based on half-integral relaxations and matroid-based kernelization, in the context of \textsc{Odd Cycle Transversal} and \textsc{Multiway Cut}.

The generic branching algorithm for \textsc{CSP 0/1/all Extension} based on half-integral relaxations turned out to be outperformed by previous approaches
to \textsc{Odd Cycle Transversal}, most notably the tailored algorithm of Akiba and Iwata~\cite{AkibaI16} based on similar principles.

For \textsc{Multiway Cut}, on sparse and planted instances the variant of the algorithm with the strongest lower bound pruning outperformed other studied approaches.
The situation changed at instances coming from the NP-hardness reduction from \textsc{Max Cut}, where the classic approach based on important separators is clearly the fastest.
However, these tests are artificial and specific. Hence, we recommend the variant denoted by \texttt{CSP LB2} as the algorithm of choice for (vertex-deletion, unweighted) \textsc{Multiway Cut}.

The preprocessing step of the \textsc{CSP 0/1/all Extension} algorithm turned out to be very powerful, resolving a significant fraction of the planted solutions
without branching and significantly reducing the sizes of sparse instances.

On the other hand, the matroid-based kernelization algorithms turned out to be impractical, requiring much more resources (both time and memory) than the solvers of the corresponding problems.

Finally, we remark that there is a need for development of more extensive and more difficult benchmark dataset for \textsc{Multiway Cut}. 
It is quite nontrivial to obtain an instance where a simple approximation algorithm gives a suboptimal solution. Our \textsc{Max Cut} instances are difficult for the algorithms
but artificial and specific. Most of the planted instances are resolved completely by the preprocessing step, while most of the sparse instances are relatively simple and quick to solve.
Furthermore, our dataset is mostly developed with evaluation of the branching algorithms in mind. Fafianie and Kratsch in their independent experiments develop a benchmark set more tailored
to evaluation of kernelization~\cite{stefans}.

\bibliographystyle{abbrv}
\bibliography{refs}

\appendix

\begin{landscape}
\begin{table}
\begin{center}
\begin{tabular}{l|lll|llll|lll|lll}
\multirow{2}{*}{test} &
\multirow{2}{*}{$|V(G)|$} &
\multirow{2}{*}{$|E(G)|$} & 
\multirow{2}{*}{$t$} &
\multirow{2}{*}{LB1} &
\multirow{2}{*}{LB2} & 
\multirow{2}{*}{OPT} &
\multirow{2}{*}{APPX} &
\multicolumn{3}{c}{reduced} &
\multicolumn{3}{c}{left} \\
&&&&&&&&
DEL & UNDEL & TERMS & $N(T)$ & OPT &  TERM \\\hline
\texttt{050\_003\_050\_30} & 200 & 3756 & 3 & 47 & 47 & 50 & 54 & 9 & 0 & 0 & 75 & 41 & 3 \\
\texttt{050\_003\_100\_20} & 350 & 6336 & 3 & 44 & 44 & 50 & 56 & 9 & 0 & 0 & 69 & 41 & 3 \\
\texttt{050\_004\_050\_20} & 250 & 3189 & 4 & 41 & 41 & 50 & 52 & 11 & 0 & 0 & 60 & 39 & 4 \\
\texttt{050\_005\_050\_20} & 300 & 3895 & 5 & 49 & 49 & 50 & 71 & 13 & 0 & 0 & 72 & 37 & 5 \\
\texttt{100\_003\_200\_20} & 700 & 25162 & 3 & 91 & 91 & 100 & 122 & 7 & 0 & 0 & 167 & 93 & 3 \\
\texttt{100\_004\_050\_30} & 300 & 8890 & 4 & 95 & 95 & 100 & 115 & 41 & 0 & 0 & 107 & 59 & 4 \\
\texttt{100\_005\_050\_30} & 350 & 10811 & 5 & 97 & 97 & 100 & 130 & 40 & 0 & 0 & 113 & 60 & 5 \\
\texttt{100\_005\_100\_20} & 600 & 15830 & 5 & 90 & 90 & 100 & 125 & 27 & 0 & 0 & 125 & 73 & 5 \\
\texttt{100\_007\_050\_20} & 450 & 9736 & 7 & 91 & 92 & 100 & 132 & 35 & 0 & 0 & 111 & 65 & 7 \\
\texttt{200\_003\_200\_20} & 800 & 40030 & 3 & 121 & 121 & 143 & 153 & 23 & 0 & 0 & 196 & 120 & 3 \\
\texttt{200\_003\_200\_30} & 800 & 59735 & 3 & 166 & 167 & 200 & 209 & 34 & 0 & 0 & 264 & 166 & 3 \\
\texttt{200\_004\_050\_40} & 400 & 25690 & 4 & 172 & 172 & 200 & 202 & 92 & 0 & 0 & 160 & 108 & 4 \\
\texttt{200\_004\_100\_20} & 600 & 24014 & 4 & 108 & 108 & 140 & 143 & 34 & 0 & 0 & 147 & 106 & 4 \\
\texttt{200\_004\_100\_30} & 600 & 35906 & 4 & 188 & 189 & 200 & 232 & 84 & 0 & 0 & 208 & 116 & 4 \\
\texttt{200\_004\_200\_20} & 1000 & 52073 & 4 & 166 & 166 & 200 & 229 & 43 & 0 & 0 & 245 & 157 & 4 \\
\texttt{200\_005\_050\_20} & 450 & 15084 & 5 & 119 & 119 & 148 & 155 & 52 & 0 & 0 & 133 & 96 & 5 \\
\texttt{200\_005\_050\_30} & 450 & 22806 & 5 & 161 & 161 & 199 & 202 & 90 & 0 & 0 & 141 & 109 & 5 \\
\texttt{200\_005\_050\_40} & 450 & 30379 & 5 & 198 & 198 & 200 & 247 & 124 & 0 & 0 & 147 & 76 & 5 \\
\texttt{200\_005\_100\_20} & 700 & 28901 & 5 & 154 & 154 & 197 & 208 & 64 & 0 & 0 & 180 & 133 & 5 \\
\texttt{200\_007\_100\_20} & 900 & 38834 & 7 & 187 & 187 & 200 & 261 & 78 & 0 & 0 & 217 & 122 & 7 \\
\texttt{200\_010\_050\_20} & 700 & 26289 & 10 & 197 & 197 & 200 & 247 & 130 & 0 & 0 & 134 & 70 & 10 \\
\end{tabular}
\caption{Results for preprocessing for \textsc{Multiway Cut} for planted instances.
    Completely resolved instances are not presented here.}\label{tb:pre-mwc:planted}
\end{center}
\end{table}
\end{landscape}

\begin{landscape}
\begin{table}
\begin{center}
\begin{tabular}{l|lll|llll|lll|lll}
\multirow{2}{*}{test} &
\multirow{2}{*}{$|V(G)|$} &
\multirow{2}{*}{$|E(G)|$} & 
\multirow{2}{*}{$t$} &
\multirow{2}{*}{LB1} &
\multirow{2}{*}{LB2} & 
\multirow{2}{*}{OPT} &
\multirow{2}{*}{APPX} &
\multicolumn{3}{c}{reduced} &
\multicolumn{3}{c}{left} \\
&&&&&&&&
DEL & UNDEL & TERMS & $N(T)$ & OPT &  TERM \\\hline
\texttt{clh-10-1\_3} & 956 & 2851 & 3 & 15 & 15 & 19 & 20 & 0 & 7 & 0 & 30 & 19 & 3 \\
\texttt{codeminer\_4} & 719 & 1006 & 4 & 10 & 10 & 10 & 11 & 5 & 83 & 1 & 9 & 5 & 3 \\
\texttt{diseasome\_0} & 1314 & 2498 & 34 & 29 & 30 & 30 & 31 & 23 & 993 & 25 & 11 & 7 & 9 \\
\texttt{fvs008\_0} & 1523 & 1835 & 263 & 107 & 108 & 111 & 122 & 84 & 684 & 227 & 45 & 27 & 36 \\
\texttt{fvs021\_0} & 1508 & 1821 & 265 & 110 & 111 & 116 & 129 & 78 & 551 & 222 & 64 & 38 & 43 \\
\texttt{fvs023\_3} & 311 & 917 & 3 & 15 & 15 & 15 & 16 & 4 & 106 & 0 & 21 & 11 & 3 \\
\texttt{fvs047\_0} & 68 & 139 & 4 & 18 & 19 & 20 & 21 & 1 & 7 & 0 & 34 & 19 & 4 \\
\texttt{fvs052\_0} & 1367 & 1912 & 140 & 187 & 188 & 189 & 207 & 173 & 850 & 121 & 27 & 16 & 19 \\
\texttt{fvs056\_0} & 1456 & 2083 & 146 & 194 & 197 & 201 & 211 & 159 & 840 & 96 & 69 & 42 & 50 \\
\texttt{fvs056\_4} & 1714 & 2384 & 4 & 11 & 11 & 11 & 14 & 1 & 564 & 0 & 19 & 10 & 4 \\
\texttt{fvs078\_0} & 2493 & 3048 & 332 & 355 & 356 & 362 & 378 & 317 & 1423 & 275 & 76 & 45 & 57 \\
\texttt{fvs100\_4} & 158 & 533 & 4 & 22 & 22 & 23 & 24 & 7 & 26 & 0 & 29 & 16 & 4 \\
\texttt{hex\_3} & 313 & 877 & 3 & 15 & 15 & 19 & 20 & 0 & 28 & 0 & 30 & 19 & 3 \\
\texttt{photoviz\_dynamic\_0} & 334 & 557 & 20 & 39 & 39 & 39 & 41 & 31 & 188 & 14 & 15 & 8 & 6 \\
\texttt{planar1500\_0} & 1407 & 3648 & 33 & 37 & 38 & 38 & 40 & 28 & 1067 & 24 & 17 & 10 & 9 \\
\texttt{planar2000\_0} & 1793 & 3592 & 79 & 78 & 79 & 79 & 87 & 74 & 1343 & 71 & 8 & 5 & 8 \\
\texttt{planar2500\_0} & 2112 & 3547 & 144 & 112 & 112 & 112 & 123 & 110 & 1423 & 141 & 3 & 2 & 3 \\
\texttt{planar3000\_0} & 2331 & 3460 & 216 & 125 & 127 & 127 & 132 & 119 & 1361 & 204 & 12 & 8 & 12 \\
\texttt{planar400\_0} & 362 & 826 & 14 & 16 & 16 & 16 & 17 & 14 & 274 & 11 & 3 & 2 & 3 \\
\texttt{planar500\_0} & 449 & 845 & 16 & 19 & 19 & 19 & 22 & 16 & 359 & 13 & 5 & 3 & 3 \\
\texttt{planar600\_0} & 502 & 809 & 41 & 30 & 30 & 30 & 32 & 28 & 317 & 38 & 3 & 2 & 3 \\
\texttt{power\_4} & 4925 & 6565 & 4 & 11 & 11 & 11 & 13 & 2 & 1495 & 0 & 17 & 9 & 4 \\
\end{tabular}
\caption{Results for preprocessing for \textsc{Multiway Cut} for sparse instances.
    Completely resolved instances are not presented here.}\label{tb:pre-mwc:sparse}
\end{center}
\end{table}
\end{landscape}

\begin{table}[tb]
\begin{center}
\begin{tabular}{l|ll|ll|ll|ll}
text & \texttt{CSP} & \texttt{+CC} & \texttt{CSPLB1} & \texttt{+CC} & \texttt{CSPLB2} & \texttt{+CC} & \texttt{ImpSeps} & \texttt{+CC} \\\hline
\texttt{maxcut\_bull} & - & - & 23:43.52 & 17:11.34 & 18:06.69 & 13:04.74 & - & 0:00.63 \\
\texttt{maxcut\_c4} & - & - & 1:10.21 & 1:51.69 & 1:00.91 & 1:36.92 & 1:38:49 & 0:00.88 \\
\texttt{maxcut\_c5} & - & - & 45:00.27 & 1:02:26 & 38:08.97 & 54:08.38 & - & 0:05.24 \\
\texttt{maxcut\_k2} & 0:00.03 & 0:00.02 & 0:00.01 & 0:00.00 & 0:00.01 & 0:00.01 & 0:00.04 & 0:00.02 \\
\texttt{maxcut\_k3} & 32:09.95 & 1:12:53 & 0:03.73 & 0:06.13 & 0:03.20 & 0:05.41 & 0:31.93 & 0:00.12 \\
\texttt{maxcut\_k4} & - & - & 20:40.45 & - & 15:34.72 & - & - & 0:06.90 \\
\texttt{maxcut\_p3} & 0:04.96 & 0:02.34 & 0:00.10 & 0:00.14 & 0:00.09 & 0:00.13 & 0:01.29 & 0:00.39 \\
\texttt{maxcut\_pan} & - & - & 15:50.32 & 16:17.25 & 13:21.68 & 13:59.79 & - & 0:01.57 \\\hline
\texttt{050\_003\_050\_30} & 0:00.15 & 0:00.16 & 0:00.08 & 0:00.07 & 0:00.08 & 0:00.07 & 0:00.07 & 0:00.07 \\
\texttt{050\_003\_100\_20} & 0:00.13 & 0:00.14 & 0:00.08 & 0:00.06 & 0:00.07 & 0:00.07 & 0:00.11 & 0:00.13 \\
\texttt{050\_004\_050\_20} & 0:08.97 & 0:09.26 & 0:00.08 & 0:00.09 & 0:00.09 & 0:00.10 & 0:00.89 & 0:01.07 \\
\texttt{050\_005\_050\_20} & 0:03.94 & 0:04.08 & 0:00.14 & 0:00.15 & 0:00.14 & 0:00.14 & 0:04.91 & 0:05.82 \\
\texttt{100\_003\_200\_20} & 0:10.46 & 0:10.97 & 0:00.75 & 0:00.77 & 0:00.75 & 0:00.77 & 0:12.04 & 0:12.73 \\
\texttt{100\_004\_050\_30} & 0:00.77 & 0:00.78 & 0:00.13 & 0:00.13 & 0:00.13 & 0:00.13 & 0:00.36 & 0:00.42 \\
\texttt{100\_005\_050\_30} & 0:02.20 & 0:02.25 & 0:00.29 & 0:00.30 & 0:00.28 & 0:00.29 & 0:00.81 & 0:00.95 \\
\texttt{100\_005\_100\_20} & 4:58.83 & 5:05.41 & 0:02.83 & 0:02.97 & 0:02.81 & 0:02.87 & 0:18.44 & 0:21.57 \\
\texttt{100\_007\_050\_20} & 1:20.03 & 1:22.92 & 0:01.17 & 0:01.21 & 0:01.14 & 0:01.19 & 0:10.52 & 0:12.45 \\
\texttt{200\_003\_200\_20} & 1:25:47 & 1:26:17 & 0:01.06 & 0:01.08 & 0:01.08 & 0:01.12 & 0:35.63 & 0:38.65 \\
\texttt{200\_003\_200\_30} & 0:19.59 & 0:19.28 & 0:03.12 & 0:03.08 & 0:03.04 & 0:03.14 & 0:06.37 & 0:06.63 \\
\texttt{200\_004\_050\_40} & 0:04.74 & 0:04.81 & 0:00.99 & 0:01.03 & 0:00.99 & 0:01.00 & 0:00.52 & 0:00.57 \\
\texttt{200\_004\_100\_20} & - & - & 0:02.34 & 0:02.48 & 0:02.30 & 0:02.37 & 0:09.25 & 0:10.71 \\
\texttt{200\_004\_100\_30} & 0:58.98 & 1:00.13 & 0:01.78 & 0:01.80 & 0:01.74 & 0:01.79 & 0:24.68 & 0:26.77 \\
\texttt{200\_004\_200\_20} & - & 1:48:33 & 0:34.71 & 0:34.20 & 0:33.70 & 0:34.86 & 48:31.29 & 50:16.20 \\
\texttt{200\_005\_050\_20} & 56:15.46 & 57:03.65 & 0:01.11 & 0:01.15 & 0:01.07 & 0:01.11 & 0:05.44 & 0:06.54 \\
\texttt{200\_005\_050\_30} & 8:15.87 & 8:27.29 & 0:04.24 & 0:04.27 & 0:04.15 & 0:04.27 & 0:05.26 & 0:06.04 \\
\texttt{200\_005\_050\_40} & 0:02.03 & 0:02.08 & 0:00.57 & 0:00.58 & 0:00.57 & 0:00.57 & 0:00.57 & 0:00.64 \\
\texttt{200\_005\_100\_20} & - & - & 0:20.22 & 0:20.78 & 0:19.93 & 0:19.90 & 1:00:45 & 1:05:37 \\
\texttt{200\_007\_100\_20} & 4:31:16 & 4:32:32 & 0:31.29 & 0:32.48 & 0:31.84 & 0:31.85 & 2:21:42 & 2:33:00 \\
\texttt{200\_010\_050\_20} & 7:47.80 & 8:08.25 & 0:03.26 & 0:03.37 & 0:03.18 & 0:03.25 & 4:05.95 & 4:51.05 \\\hline
\texttt{clh-10-1\_3} & - & 2:37:11 & 0:00.05 & 0:00.05 & 0:00.05 & 0:00.05 & 0:01.45 & 0:01.87 \\
\texttt{fvs008\_0} & 0:00.87 & 0:01.36 & 0:00.14 & 0:00.27 & 0:00.14 & 0:00.25 & - & - \\
\texttt{fvs021\_0} & 0:05.93 & 0:11.78 & 0:00.41 & 0:00.64 & 0:00.36 & 0:00.58 & - & - \\
\texttt{fvs052\_0} & 0:00.05 & 0:00.03 & 0:00.04 & 0:00.02 & 0:00.02 & 0:00.02 & - & - \\
\texttt{fvs056\_0} & 0:33.03 & 0:00.74 & 0:05.55 & 0:00.17 & 0:05.42 & 0:00.17 & - & - \\
\texttt{fvs056\_4} & 0:00.25 & 0:00.86 & 0:00.01 & 0:00.02 & 0:00.01 & 0:00.01 & 0:00.75 & 0:01.29 \\
\texttt{fvs078\_0} & 18:15.31 & 33:35.99 & 1:07.25 & 1:52.67 & 0:57.98 & 1:37.43 & - & - \\
\texttt{fvs100\_4} & 0:00.04 & 0:00.04 & 0:00.01 & 0:00.01 & 0:00.01 & 0:00.01 & 0:00.23 & 0:00.09 \\
\texttt{hex\_3} & - & - & 0:00.19 & 0:00.19 & 0:00.18 & 0:00.19 & - & - \\
\texttt{planar1500\_0} & 0:00.02 & 0:00.01 & 0:00.02 & 0:00.01 & 0:00.01 & 0:00.01 & 0:06.48 & 0:00.08 \\
\texttt{planar2000\_0} & 0:00.02 & 0:00.01 & 0:00.02 & 0:00.01 & 0:00.01 & 0:00.02 & 34:37.19 & 0:00.05 \\
\texttt{planar2500\_0} & 0:00.02 & 0:00.01 & 0:00.02 & 0:00.02 & 0:00.02 & 0:00.02 & - & 0:00.93 \\
\texttt{planar3000\_0} & 0:00.03 & 0:00.02 & 0:00.02 & 0:00.02 & 0:00.02 & 0:00.01 & - & 0:08.43 \\
\texttt{power\_4} & 0:00.02 & 0:00.02 & 0:00.02 & 0:00.02 & 0:00.02 & 0:00.02 & 0:55.77 & 0:28.37 \\
\end{tabular}
\caption{Running times for \textsc{Multiway Cut}. The second section corresponds to planted instances and the last to sparse.
  A number of sparse tests with running times below half a second for all algorithms have been omitted.
  Dash means exceeded timeout of 5 hours.}\label{tb:mwc}
\end{center}
\end{table}

\end{document}